\documentclass{article}
\usepackage[utf8]{inputenc}
\usepackage{listings}
\usepackage{url}
\usepackage{geometry}
\usepackage{color}
\usepackage{hyperref}
\usepackage{easy-todo}
\usepackage{graphicx}
\usepackage[dvipsnames]{xcolor}
\usepackage{mathtools}

\begin{document}
\vspace*{0.2in}

\begin{flushleft}
{\Large
\textbf\newline{Brain-inspired Distributed Cognitive Architecture}
}
\newline

Leendert A Remmelzwaal \textsuperscript{1*},
Amit K Mishra \textsuperscript{1},
George F R Ellis \textsuperscript{2}
\\
\bigskip
\textbf{1} Department of Electrical Engineering, University of Cape Town, Rondebosch, Cape Town,
South Africa 7700.
\\
\textbf{2} Department of Mathematics and Applied Mathematics, University of Cape Town, Rondebosch, Cape Town, South Africa 7700.
\\
\bigskip

* Corresponding author: leenremm@gmail.com

\end{flushleft}


\begin{abstract}
In this paper we present a brain-inspired cognitive architecture that incorporates sensory processing, classification, contextual prediction, and emotional tagging. The cognitive architecture is implemented as three modular web-servers, meaning that it can be deployed centrally or across a network for servers. The experiments reveal two distinct operations of behaviour, namely high- and low-salience modes of operations, which closely model attention in the brain. In addition to modelling the cortex, we have demonstrated that a bio-inspired architecture introduced processing efficiencies. The software has been published as an open source platform, and can be easily extended by future research teams. This research lays the foundations for bio-realistic attention direction and sensory selection, and we believe that it is a key step towards achieving a bio-realistic artificial intelligent system.\\


\end{abstract}


\section{Introduction}

In this paper we attempt to answer the question: Can we design a cognitive architecture based on the modular structure and functionality of the human brain? Cortical modules such as the corticothalamic circuitry \cite{alitto2003corticothalamic}, the ascending arousal systems \cite{edelman2007learning} and vision \cite{wilson1992log} are the result of of millions of years of Darwinian natural selection and survival-driven optimisation \cite{ellis2018top} \cite{ellis2017beyond}; in this research we allow these structures to guide the design of our cognitive architecture. 
We present a proof-of-concept of a distributed, computationally efficient, brain-inspired cognitive architecture that has the ability to tag events with salience and direct cortical attention to the most salient current event. In addition, this framework is open-source and can be easily extended by future research teams.


\section{Motivation}

Why did we embark on this project? It is common knowledge that bio-realistic artificial intelligence is still a distant dream to many researchers. We embarked on this work because we believe that a modular bio-inspired cognitive architecture could be a key step in designing a bio-realistic artificially intelligent system. This is a radically different approach to taking existing software (e.g. OpenCV) and adapting it to be more brain-like - it is fundamentally brain-inspired. 
It be one of the key steps towards achieving artificial intelligence. 

A second motivation was to build a model to simulate some of the behaviours of key modules in the human brain, including the thalamus, cortex, arousal system, basal ganglia, and amygdala, to help medical practitioners understand the effects of disease or accidental damage to the cortex. Wouldn't it be great if we could simulate the effect of removing the thalamus, impeding the release of noradrenaline, or triggering a seizure without relying on living test subjects? A more extended model of the kind we present could help here.

A third motivation was to design a cognitive architecture with future research in mind. The distributed modular nature of the architecture we present is designed to be extended, whereby a research group can independently enhance a single module in the system, such as the arousal system, without affecting the working of the entire system. This would allow multiple research teams to incrementally enhance the entire system by focusing on only a part of the system.

A fourth motivation was to design a scaleable infrastructure that could be deployed on a distributed network of servers, allowing the system to exist outside of just a single server. The implementation leverages a distributed network of servers, giving autonomy to computational units inspired by key units in the human brain. While this allows the system to scale, it also allows the system as a whole to build in redundancies that will improve efficiency and robustness of the system as a whole.

Finally, we hope that building on the research presented in this paper will eventually shed light on the nature of consciousness \cite{solmsfriston2018consciousnessarises} \cite{solms2019hardproblem}.


\section{Key Structures in the Brain}

\begin{figure}[!h]
  \centering
  \includegraphics[scale=0.5]{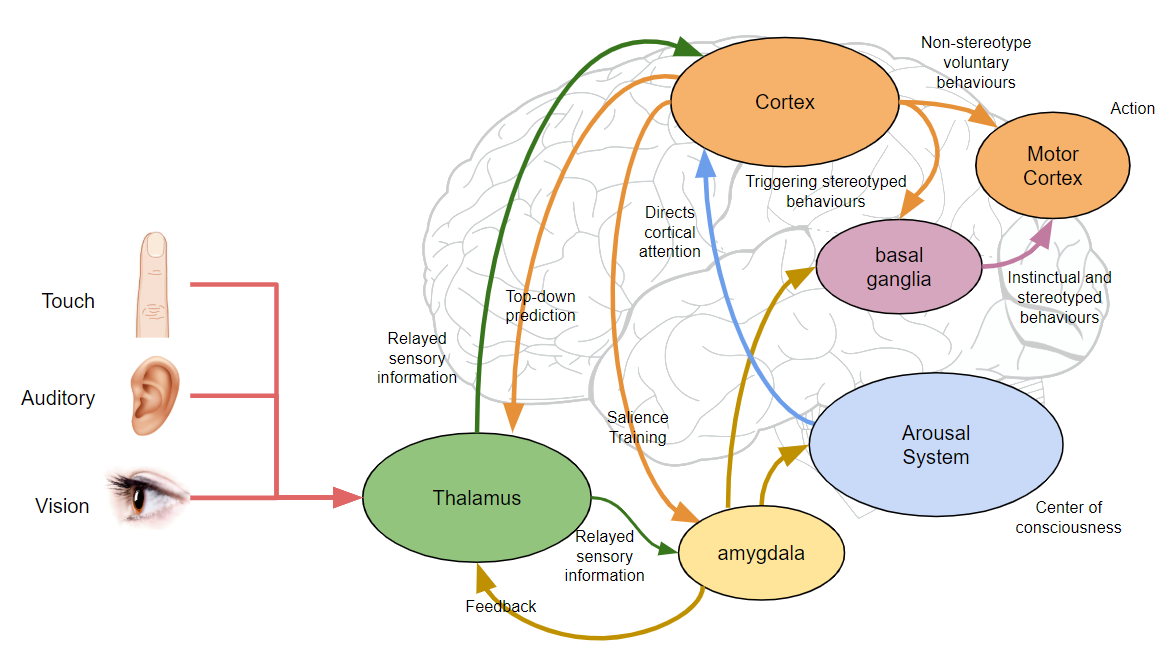}
  \caption{Some of the key autonomic and voluntary pathways between the thalamus, amygdala, arousal system, basal ganglia and the prefrontal cortex.}
  \label{fig:Fig001}
\end{figure}

Our bio-inspired cognitive architecture is modelled closely after a number of key structures in the human brain, namely the thalamus, cortex, arousal system, basal ganglia, and amygdala (see Fig \ref{fig:Fig001}). These structures play important roles in sensory processing, prediction, emotional tagging, attention direction and consciousness \cite{ellis2005neural} \cite{solms2019friston}. We model some aspects of these interactions.

\subsection{Arousal System}

Emotional tagging of events and memories is key component to consciousness, intelligence and directing attention in the cortex \cite{damasio1995descartes} \cite{panksepp2004affective} \cite{panksepp2012archaeology} \cite{solms2019friston}. The arousal system releases neuromodulators such as dopamine and noradrenaline into the cortex via diffuse projections \cite{edelman1987neural} \cite{edelman2004wider} \cite{panksepp2004affective}. These long-range diffuse projections allow neuromodulators to impact an entire pattern of synaptic connections proportional to their synaptic activation level and the strength of the neuromodulator released. The release of noradrenaline has a direct impact of increasing gain on perception, which in turn also directs cortical attention to the object or event causing the emotional response. For example, if someone has a fear of dogs, then seeing a dog may make them feel afraid, which in turn commands their attention to the dog. In this paper we model the way (a) the arousal system releases noradrenaline, and (b) how the noradrenaline level decays over time.

\subsection{Thalamus}

The thalamus acts as a relay station between visual, auditory and somatosensory sensory inputs and the prefrontal cortex \cite{alitto2003corticothalamic} \cite{miller2002spectrotemporal} \cite{sillito2002corticothalamic} \cite{briggs2008emerging}. The connections between the thalamus and the prefrontal cortex are referred to as corticothalamic connections \cite{sherman2009exploring} and consist of both bottom-up (feed forward) and top-down (feedback) connections \cite{ellis2018top}. The thalamus is responsible for comparing the incoming signal to the top-down prediction signal, and sending a moderated signal back to the cortex if the difference between the predicted signal and incoming data exceeds a threshold. This threshold can be affected by the level of neuromodulators released from the arousal system, which is a key feature of our model. In this paper we focus on modelling the way the thalamus (a) relays incoming information to the cortex, (b) receives a contextual prediction from the cortex, and (c) compares the incoming information to the prediction before deciding whether to send the next image to the cortex.

\subsection{Prefrontal Cortex}

These prefrontal cortex is responsible for a range of key functions such as cognition \cite{jones2002thalamic}, attention direction \cite{shipp2004brain} \cite{de2014thalamic} \cite{wimmer2015thalamic} \cite{friston2018does}, sensory selection \cite{thalamic_saalmann2014neura_attentionl} \cite{thalamic_control_attention} \cite{thalamic_zikopoulos2007circuits}, awareness \cite{rees2009visual}, emotional control \cite{sun2015human}, prediction \cite{ellis2017beyond}, and salience detection \cite{bowman2013attention}. In this paper we focus on modelling the way the cortex (a) receives sensory inputs, (b) classifies the sensory input, (c) creates a prediction of the current context, (d) produces a desire to act, and (e) triggers the release of neuromodulators. The model is a simplified integrated  representation of these features.


\section{Scope of Work}

In summary: we present a proof-of-concept of a brain-inspired cognitive architecture that incorporates sensory processing, classification, contextual prediction, and emotional tagging. The architecture is robust and modular; meaning that it can be deployed centrally or across a network for servers. The software produced by this research has been published as an open source platform, and can be easily extended by future research teams.


\section{Related Work}

Our work builds on existing research by connecting together models of individual modules in the brain, namely models of the thalamus, cortex, arousal system.

In modelling the thalamus, cortex and corticothalamic connections, we reviewed the work done on predictive processing by Andy Clark \cite{clark2015embodied}, as well as the models of Retinal Predictive Coding \cite{srinivasan1982predictive}, Linear Predictive Coding \cite{o1988linear}, Cortical Predictive Coding \cite{rao1999predictive}, Restricted Boltzmann machine (RBM) \cite{hinton2006fast}, Free Energy Predictive Coding \cite{friston2009predictive}, BC-DIM Predictive Coding \cite{spratling2016predictive}, Predictive Sparse Decomposition \cite{kavukcuoglu2010fast}, Stacked Denoising Auto-encoders \cite{vincent2010stacked}, Deep Predictive Coding Networks \cite{chalasani2013deep}, PredNet \cite{lotter2016deep}, Multilevel Predictor Estimator \cite{kim2017predictor}, Deep Predictive Coding \cite{wen2018deep}, LPCNet \cite{valin2019lpcnet}, and the CTNN model \cite{remmelzwaal2019ctnn}. We have chosen to incorporate the CTNN model because unlike the other models, it is input agnostic, multi-modal and computationally efficient.

In modelling the arousal system, we reviewed the work done on modelling non-local effects of neuromodulators by Edelman and the primary emotional systems identified by Panksepp \cite{panksepp2004affective}. These models include the Husbands' model of gas diffusion \cite{husbands1998evolving}, Juvina's model of valuation and arousal\cite{juvina2018modeling}, Edelman's brain-based devices \cite{edelman2007learning} \cite{krichmar2005brain}, and the Salience-affected Neural Network (SANN) \cite{remmelzwaal2019sann}. We have chosen to use the SANN architecture to model the effects of salience because unlike the other models salience affects a specific neural activation pattern as a whole, tagging it with a salience signal, and with one-time learning.

Attempts have previously been made to map the connectome, such as the Blue Brain Project \cite{markram2006bluebrainproject}, but we are focused on the functional structures of each module in the brain, rather than the connections between individual neurons. Our research is therefore at a level abstracted from mapping the connectome.


\section{Implementation}

\begin{figure}
  \centering
  \includegraphics[scale=0.65]{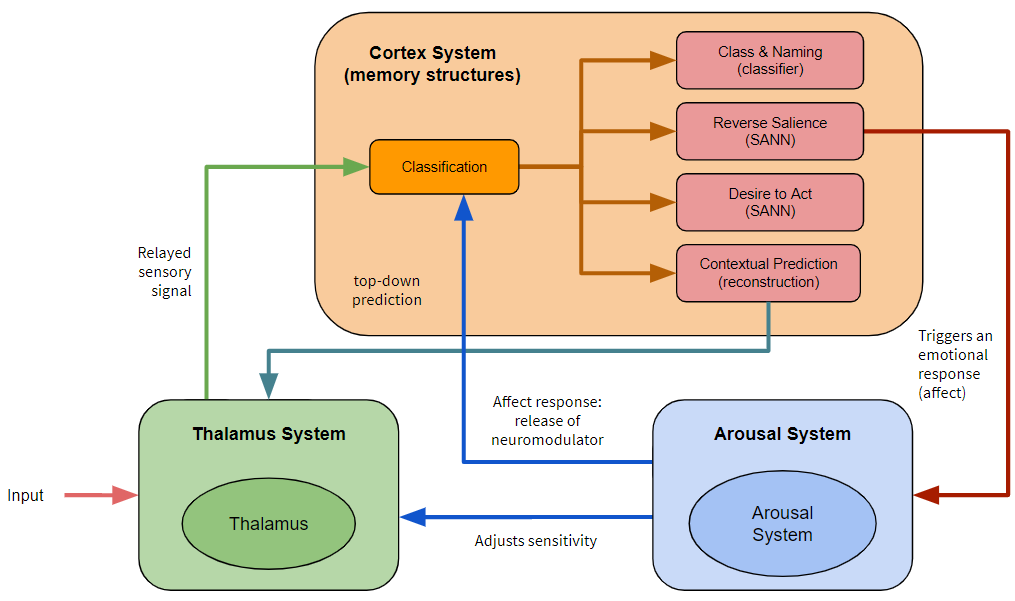}
  \caption{Overview of the distributed system architecture.}
  \label{fig:Fig003}
\end{figure}

The distributed computational model we use is represented in Fig \ref{fig:Fig003}, developed for computational purposes from the set of interactions summarised in  Fig 1. The thalamus, arousal system and memory structures were each modelled as their own independent Python Flask web-servers. Each web-server could be called by any other web-server, to request information (e.g. current state of the cortex) or to request an action (e.g. classifying a new image). For more technical information on the API protocols see \cite{remmelzwaal2020python}.

\subsection{Modelling the Arousal System}

The arousal system was modelled as the first stand-alone server. It maintains as an internal state the levels of neuromodulators such as noradrenaline in the cortex. Any other system (e.g. the cortex) may trigger the release of a neuromodulator, and may request the current level of neuromodulator.

Once a neoromodulator has been released, it is programmed to decay at a fixed rate as represented in (1).
Here $x$ is the number of seconds that has passed.

\begin{equation}
y = e^{-0.2 * x}
\end{equation}

\subsection{Modelling the Thalamus}

The thalamus system was modelled as the second stand-alone web-server. It calculates an internal threshold depending on the level of noradrenaline released by the arousal system. If the difference between the incoming image and the contextual prediction generated by the cortex system exceeds the threshold, the thalamus sends the input image to the cortex system for processing. A key point here is that the thalamus threshold is directly impacted by the level of noradrenaline release by the arousal system. The higher the level of noradrenaline, the lower the internal threshold in the thalamus system. This models the work of Solms and Friston on the relation between arousal and conscousness.  

\subsection{Modelling the Cortex}

The cortex system was modelled as the third stand-alone web-server, and incorporates the SANN \cite{remmelzwaal2019sann} and CTNN models previously published by Remmelzwaal et al \cite{remmelzwaal2019ctnn}. The cortex system accepts an input image, and processes this input image using a number of internal models, including an auto-encoder (based on the CTNN) to reduce image dimensionality and to generate a reconstruction of the current context, and deep neural network (based on the SANN) that classifies and returns the salience tag for each image (see Fig \ref{fig:Fig003}).

\begin{figure}[!h]
  \centering
  \includegraphics[scale=0.7]{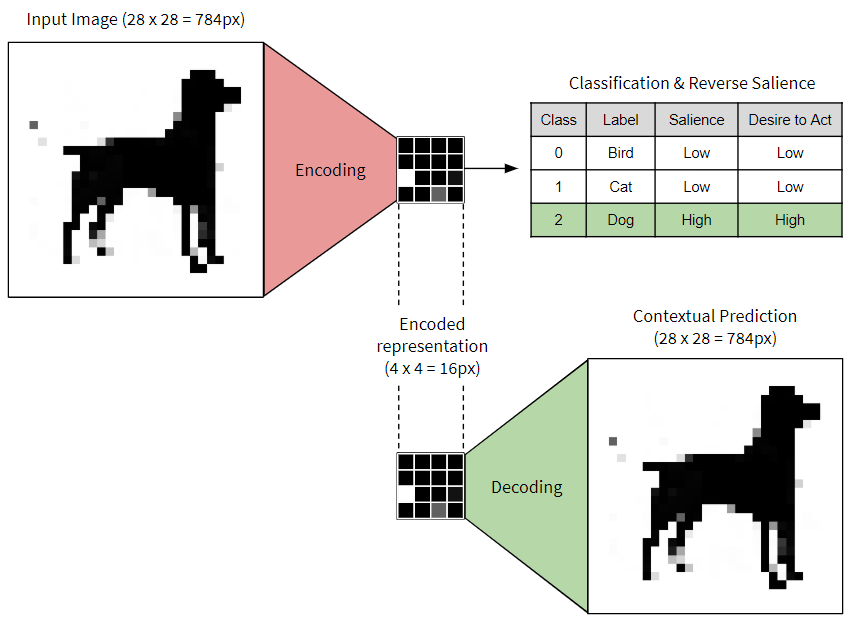}
  \caption{Illustration of information flow in the cortex system.}
  \label{fig:Fig004}
\end{figure}

\subsection{Graphical User Interface (GUI)}

\begin{figure}[!h]
  \centering
  \includegraphics[scale=0.5]{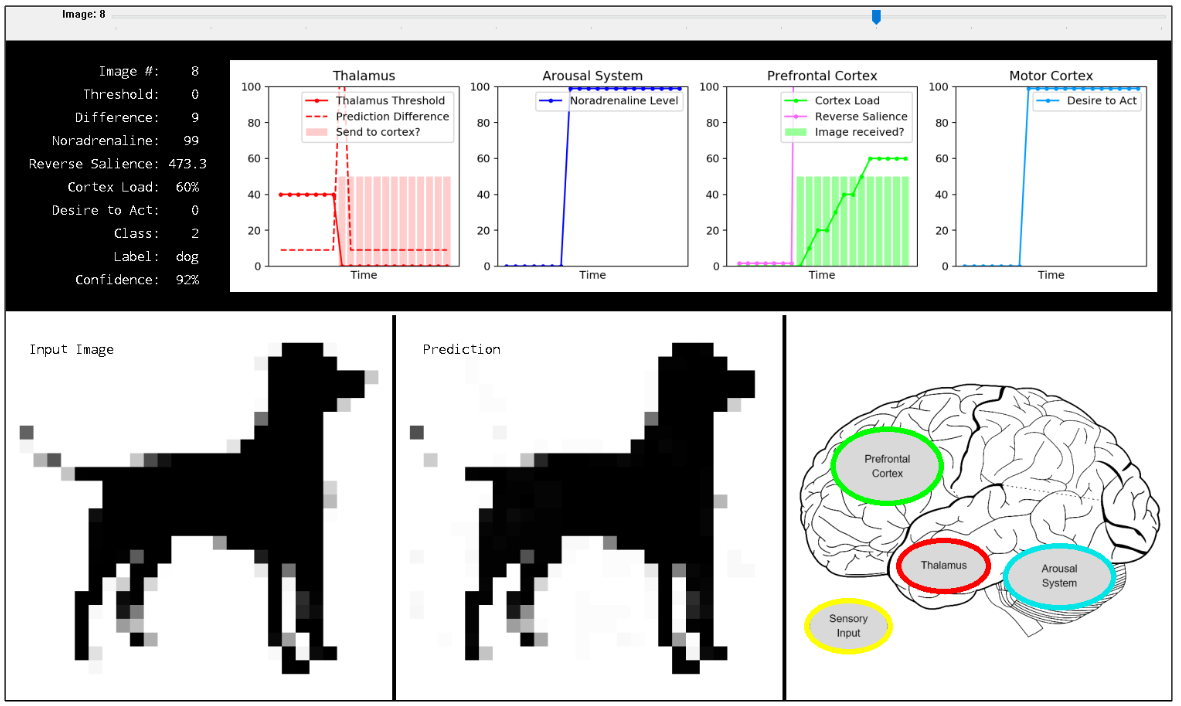}
  \caption{The Graphical User Interface (GUI)}
  \label{fig:Fig005}
\end{figure}

A graphical user interface (GUI) shown in Fig \ref{fig:Fig005} was created to allow the researcher to vary the input image, and to give the researcher visibility of key variables in the system (e.g. thalamus threshold, cortical prediction, neuromodulator level).

\section{Dataset}

In this paper we generated a dataset of black and white silhouettes of animals in three classes, namely Bird, Cat and Dog (shown in Fig \ref{fig:Fig001}). This dataset was inspired by \cite{kaggle-animals10}, and is a novel dataset presented for the first time here.


\section{Training}

In this model only the memory structures in the Cortical System are trainable. Via the GUI or API interface the researcher can initiate training of both the auto-encoder (used to predict context) and the SANN (to classify and determine salience). For the purposes of demonstration, a single image of a DOG was tagged with one-time salience training.


\section{Results and Discussion}

The experiments conducted demonstrate various key results.

\subsection{Low- and high-salience modes of operation}

Firstly, the experiments reveal two distinct operations of behaviour, which closely model attention in the brain.

\begin{figure}[!h]
  \centering
  \includegraphics[scale=0.6]{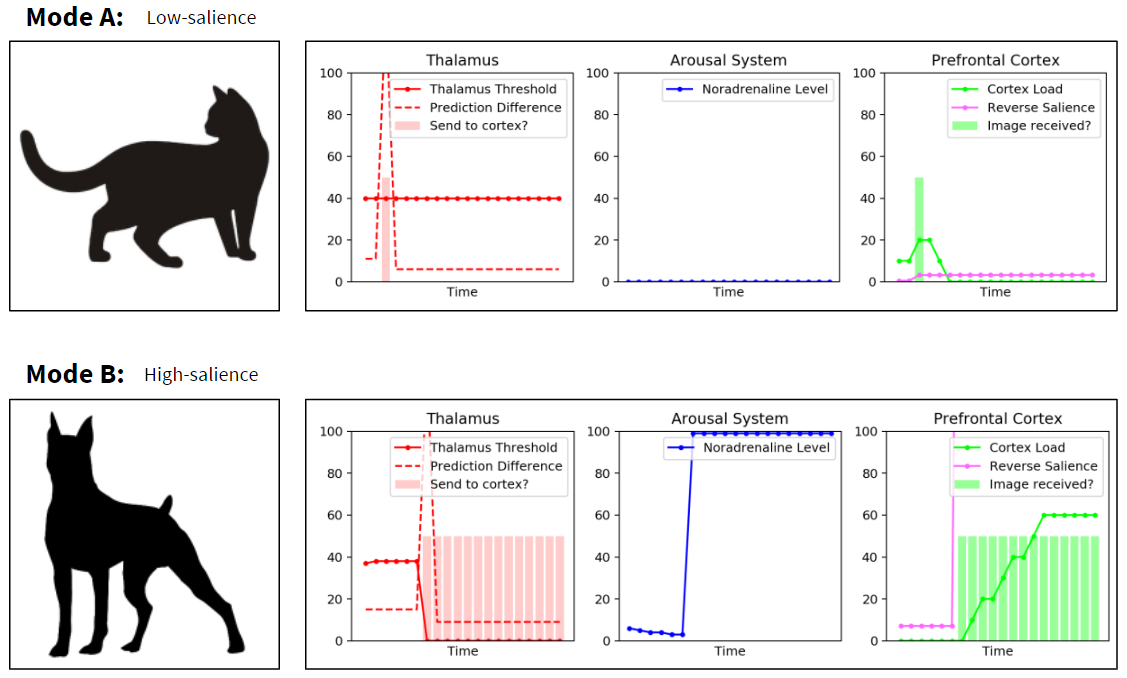}
  \caption{Two modes of operation. Low-salience (A) and High-salience (B) inputs}
  \label{fig:Fig007}
\end{figure}

The first mode of operation occurs when the system is presented with an input image that has a low-salience association (see Fig \ref{fig:Fig007}A). When initially presented with such an image, the thalamus notices a significant change in input, and engages the cortex to classify the image. The cortex is engaged for one computational iteration, and returns a contextual prediction to the thalamus. For subsequent viewings with the same image the thalamus does not engage the cortex. During this mode, the cortex directs attention momentarily to identify an object, but then the cortex is free to assign attention to another classification or prediction task.

The second mode of operation occurs when the system experiences an input image with a high associated salience (see Fig \ref{fig:Fig007}B). Similar to the first mode of operation, the thalamus notices a significant change in input, and engages the cortex to classify the image. Again, the cortex is engaged for one computational iteration, and returns a contextual prediction to the thalamus. However this time, the cortex identifies a high salience association with the high level concept, and two things happen: firstly, the cortex triggers a release of noradrenaline from the arousal system, and secondly, the release of noradrenaline lowers the internal threshold of the thalamus System. This has a direct impact on how the system processes subsequent images. With a lower internal threshold, the thalamus System sends images to the cortex more frequently, demanding the computational power and attention of the cortex system. In this mode, the cortex directs all of its attention to the incoming images, ignoring other lower-priority tasks.

\subsection{Confirming neuroscience understanding}

Secondly, the simulations conducted in this paper confirm neuroscience understanding: that the thalamus and arousal system indeed play a crucial roles in processing incoming data, and directing cortical attention. Without the Thalamus constantly comparing incoming signals to a contextual prediction, our brains would be overwhelmed with a non-stop overload of incoming information to process. Without salience tagging of memories and event enabled by the Arousal system, the cortex wouldn't receive the guidance it requires to direct its attention to the most salient issue

\subsection{Optimizing performance of complex tasks}

In addition to modelling the thalamus and arousal system, we have also demonstrated that a bio-inspired architecture can optimize performance of complex processing tasks (e.g. classification and prediction). We have show that our architecture frees up computational processing during low-salience events, especially when there is high levels of similarity between one image an another. We experience this in the physical world almost every day when we drive - driving at a steady speed down a road is a highly monotonous task, whereby your brain is free to engage in conversations or to sing along to a song on the radio. However, if a person steps onto the road, your attention is immediately directed to managing the salience situation, abandoning almost instantly the conversation you were having, or song you were singing. We have demonstrated this effect by adding 2 sub-systems inspired by the functions of the the thalamus and the arousal system in the brain.

\subsection{Attention, salience and prediction}

In this paper we have demonstrated with the aid of a fully-connected simulation of the thalamus, arousal system and cortex, how these three key structures rely on each other to allow for:

\begin{enumerate}

  \item Attention direction \cite{shipp2004brain} \cite{de2014thalamic} \cite{wimmer2015thalamic} \cite{friston2018does}; in our model the cortex is engaged with only those images that differ significantly from the predicted context, a level which is modulated by the affective response from previous images. This has the impact of freeing up the cortex for other processing during low-salience events (e.g. driving a car), and demands attention from the cortex during high-salience events (e.g. seeing a dog). What is important to observe is that in our model the cortex is not actively directing its own attention; it is a emergent feature guided by the arousal system and the thalamus.
  
  \item Salience detection \cite{bowman2013attention}; our system allows for events to be tagged with affect (salience), and for that salience to be returned along with the label if that combination of sensory inputs is experiences again in the future.
 
  \item A simple form of prediction \cite{ellis2017beyond}; the cortex 
  generates a prediction of the current context by leveraging an auto-encoder. The model sends this   prediction of the current context back to the thalamus system to be compared with incoming sensory information. 
  \end{enumerate}

\subsection{Extensibility}

We have designed a cognitive architecture specifically with future research in mind. The distributed modular nature of the architecture we present here allows research groups to  independently enhance and extend a single module in the system (e.g. the arousal system) without affecting the functioning of the entire system. The inner working of each module is unknown to other modules, and each module can only interact with other modules using predefined API calls. For example, the thalamus can send the cortex a new sensory image, and can request a prediction, but it is not concerned with how the cortex processes the information it is sent, or generates the prediction. We believe that this approach to designing the cognitive architecture is key in allowing this research to be extended in the future.


\section{Future Work}

In this paper we have presented an open-source brain-inspired distributed cognitive architecture. The Graphical User Interface (GUI) allows researchers to examine and interact with the model. We hope that future cognitive architectures can benefit from the principles presented in this paper. 

This model can be extended to include recurrent connections, or to include prediction over a period of time (e.g. to simulate autonomous behaviour when driving a car). While the current model accepts only images, it can be extended to multi-modal inputs such as auditory of somatosensory inputs, as is demonstrated in \cite{remmelzwaal2019ctnn}. It may also be valuable to explore whether the system proposed in this paper experiences subjective states, using an evaluation framework such as the Independent Core Observer Model (ICOM) \cite{kelley2019measures}.\\

\noindent{}We believe that this system could be extended to allow for:
\begin{enumerate}
  \item Emotional control \cite{sun2015human}; how does an emotional response translate to an action? that can guide behaviour of the system. This would extend the "desire to act" feature of our current system.
  
  \item Sensory selection \cite{thalamic_saalmann2014neura_attentionl} \cite{thalamic_control_attention} \cite{thalamic_zikopoulos2007circuits}; how can the cortex select to  attention from a specific sensory input? This would require extending the thalamus model to allow sending of multiple modes of sensory data, and for the cortex to be able to process multiple modes of sensory data independently of each other.
\end{enumerate}
  The ultimate aim of this paper has been to introduce a cognitive architecture that may help lead to better modelling and understanding of  awareness \cite{rees2009visual}, consciousness \cite{solmsfriston2018consciousnessarises}, and subjective experiences \cite{kelley2018subjectiveexperience}. We believe the design principles introduced here can be important in that search.


\section{Supporting Material}

The source code as well as records of the tests conducted in this paper are publicly available online \cite{remmelzwaal2020python}. For additional information, please contact the corresponding author.


\section{Acknowledgements}

\noindent{This research did not receive any specific grant from funding agencies in the public, commercial, or not-for-profit sectors.}


\end{document}